\documentclass[pra,aps,twocolumn,longbibliography]{revtex4-1}
\usepackage{graphicx,multirow}
\usepackage{color,outlines}
\usepackage[sc,osf]{mathpazo}
\usepackage{amsmath,amssymb}
\usepackage{thmtools}

\usepackage{physics}
\usepackage{outlines}
\usepackage{tikz}
\usetikzlibrary{positioning}
\usetikzlibrary{arrows}
\usepgflibrary{arrows}
\usetikzlibrary{arrows.meta,automata, decorations.pathreplacing}

\usepackage{hyperref}
\usepackage{graphicx}
\usepackage{dcolumn}
\usepackage{bm}
\usepackage{color}
\usepackage{xcolor}
\usepackage{bbold}
\usepackage{soul}
\usepackage{epsfig,epic}
\usepackage{epstopdf}
\usepackage{amsthm}
\usetikzlibrary{decorations.text,calc,arrows.meta}

\usepackage{cleveref}
\usepackage{algorithm2e}

\newcommand{\newc}{\newcommand}
\newc{\beq}{\begin{equation}}
\newc{\eeq}{\end{equation}}
\newc{\kt}{\rangle}
\newc{\br}{\langle}
\newc{\beqa}{\begin{eqnarray}}
\newc{\eeqa}{\end{eqnarray}}
\newc{\ovl}{\overline}
\usepackage[flushleft]{threeparttable} 
\usepackage{tabularx}
\usepackage{array}
\usepackage{soul}

\begin{document}

\title{Dual unitaries as maximizers of the distance to local product gates} 
\author{Shrigyan Brahmachari}
\thanks{ sb818@duke.edu\\
Current address: Department of Electrical \& Computer Engineering,
Duke University Pratt School of Engineering,
Box 90291 Durham, NC 27708 USA.} 
\affiliation{Department of Mechanical Engineering, Indian Institute of Technology
Madras, Chennai, India~600036}
\author{Rohan Narayan Rajmohan}
\thanks{Current address: Department of Physics and Astronomy, Northwestern University, Evanston, IL 60208, USA} 
\author{Suhail Ahmad Rather}
\thanks{Current address: Max Planck Institute for the Physics of Complex Systems, N\"othnitzer Str. 38, 01187 Dresden, Germany.}
\author{Arul
Lakshminarayan}
\thanks{arul@physics.iitm.ac.in} 
\affiliation{Department of Physics, Indian Institute of Technology
Madras, Chennai, India~600036}
   
\begin{abstract}
The problem of finding the resource free, closest local unitary, to any bipartite unitary gate $U$ is addressed. Previously discussed as a measure of nonlocality, the distance $K_D(U)$ to the nearest product unitary has  implications for circuit complexity and related quantities. Dual unitaries, currently of great interest in models of complex quantum many-body systems, are shown to have a preferred role as these are maximally and equally away from the set of local unitaries. This is proved here for the case of qubits and we present strong numerical and analytical evidence that it is true in general. An analytical evaluation of $K_D(U)$ is presented for general two-qubit gates. For arbitrary local dimensions, that $K_D(U)$ is largest for dual unitaries, is substantiated by its analytical evaluations for an important family of dual-unitary and for certain non-dual gates. A closely allied result concerns, for any bipartite unitary, the existence of a pair  of maximally entangled states that it connects. We give efficient numerical algorithms to find such states and to find $K_D(U)$ in general.
\end{abstract}

\maketitle

Dual-unitary quantum circuits are of intense current interest to many research communities \cite{Akila_2016,Sarang_2019,Bertini_2018,Bertini_2019,Gutkin_2020,Claeys_2020,Lorenzo_2020,Claeys_2021,ASA_2021,Alessio_2021,Tianci_2022} as they provide non-trivial models of both integrable and chaotic many-body quantum systems and allow for universal computation. For example, these have been used for the evaluation of dynamical correlation functions \cite{Bertini_2019,Gutkin_2020}, spectral statistics \cite{Bertini_2018}, construction of a quantum ergodic hierarchy \cite{ASA_2021}, entanglement generation \cite{Lorenzo_2020}, exact emergence of random matrix universality \cite{Choi_2022}, and measurement induced phase-transitions \cite{Claeys_2022}. It has been shown to be classically simulatable for short times or circuit depths for certain initial states and for local expectation values \cite{Ryotaro_2022}. However, for late times the problem has been shown to be BQP-complete and the dual-unitary circuits are capable of universal quantum computation, while classical simulation of the problem of sampling has been shown to be hard \cite{Ryotaro_2022}. 

The building blocks of dual-unitary circuits are arbitrary single-particle gates and two-particle dual unitary operators. The dual-unitary operators remain unitary on reshuffling the indices, a property interpreted as a space-time duality  \cite{Bertini_2019}. 
From a quantum information theoretic viewpoint these are maximally entangled unitaries \cite{SAA_2020,Tyson_2003}. Nevertheless, their place in the space of general bipartite unitary operators, $\mathbb{U}(d^2)$ in local dimension $d$, is not understood and their construction for $d>2$ is incomplete \cite{Bertini_2019}. However, there are numerical algorithms \cite{SAA_2020} to generate ensembles of dual-unitary matrices in any dimension $d$ and several analytic constructions \cite{Gutkin_2020,Claeys_2021,ASA_2021,Borsi_2022}. 

As local unitary operators are considered to be a free resource, a basic geometric question is the distance of dual unitaries and in general any bipartite unitary operator to the closest local unitary. This has been previously studied as a ``strength measure" of bipartite unitary operators and denoted as $K_D(U)$ \cite{Nielsen2003}, and satisfies the conditions required of a quantum complexity measure. Formally, for a general metric $D$, $K_D(U)$ for some bipartite unitary $U \in \mathbb{U}(d^2) $ is defined as, 
\begin{equation}
    K_D(U):=\min_{u_A,u_B \in \mathbb{U}(d)}D(U,u_A \otimes u_B).
    \label{eq:kd_def}
\end{equation}

\begin{figure}[h]
\begin{center}
\begin{tikzpicture} 

  \draw(0,-9)[fill=cyan!40] circle ( 3 cm);
  \node[draw=none,align=center,text width = 8.25cm] at (0,-7.1) {all unitary operators}; 
  \draw[fill=white!40](-0,-9) circle (1 cm);
  \node[draw=none,align=center, text width = 2.25cm] at (0,-9.4) {local unitary products};
  \draw[decoration={text along path,reverse path,text align={align=center},text={Dual Unitaries}},decorate] (3,-9.3) arc (0:180:3);
\draw[black, thick] (1.75,-7.5) -- (.75,-8.25);
 \filldraw[red] (1.75,-7.5) circle (2pt) node[black,anchor=west]{U};
 \draw[black, thick,decorate,decoration={brace,amplitude=10pt}] (1.75,-7.5) -- (.75,-8.25);
 \node at (1.925,-8.25) {$K_D(U)$};
  \filldraw[blue] (.75,-8.25) circle (2pt); 
 \node at (.3,-8.4) {$u_1 \otimes u_2$};
 \filldraw[blue] (.75,-9.75) circle (2pt);
 \filldraw[red] (2.13,-11.13) circle (2pt);
 \draw[black, thick,decorate,decoration={brace,amplitude=1 0pt}] (.75,-9.75) -- (2.13,-11.13);
 \draw[black, thick] (.75,-9.75) -- (2.13,-11.13);
 \node at (1.58,-9.9) {$K_D^{*}$};
 

\end{tikzpicture} 
\caption{A caricature of the geometry of $K_D(U)$. For a given bipartite unitary operator, the projection to the subset of local unitary products is found, and the distance is calculated.}
\label{fig:cartoon}
\end{center}
\end{figure}
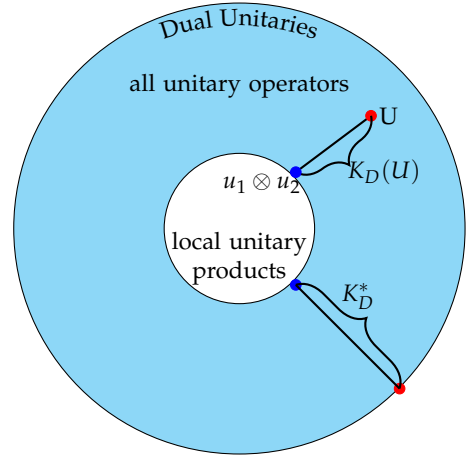
This work uses the Hilbert-Schmidt metric $\|A\|=\sqrt{\tr(A A^{\dagger})}$, as it seems both most appropriate and accessible to analytical considerations.
As a strength measure, one can choose the metric to satisfy desirable properties based on the application \cite{Nielsen2003}. It should lead to a function on unitaries that satisfy: (i) $f(u_A \otimes u_B)=0$, (ii) $f(UV) \leq f(U)+f(V)$ and (iii) $f(U \otimes \mathbb{I})=f(U)$. The first two properties are satisfied by all strength measures, and we show that if we pick the Hilbert-Schmidt metric, the third property, known as stability, is satisfied by $K_D(U)$, see Appendix~\ref{app:stability}. A particular application of interest is in circuit complexity; one can show that the strength measure on the operator norm  of a bipartite gate  can be used to understand its capacity to create entanglement when acting upon pure states, based on recent arguments in Ref. \cite{Eisert_2021}.

Thus far there are partial results concerning $K_D(U)$ for $d=2$ and no known results for $d>2$. Apart from deriving several new results concerning general bounds and exact evaluations of this measure, we show that the set of dual-unitaries are {\em maximally} and {\em equally} away from product unitaries for the case of qubits, $d=2$, and present strong numerical evidence that this is the case even for $d>2$. This motivated the caricature in Fig.~(\ref{fig:cartoon}) which indicates the special place of dual-unitaries in the space of bipartite operators.

These considerations are intimately related to an interesting property that is easy to see for qubits, but also appears to hold in general: for any bipartite unitary $U$  in $\mathcal{H}^d\otimes \mathcal{H}^d$, there exists a pair of maximally entangled states $|\Phi_1\rangle$ and $|\Phi_2\rangle$ such that $U |\Phi_1\rangle=|\Phi_2\rangle$.  A dynamical map is devised that converges to such a pair of maximally entangled states for any $U$, which can be used to find the closest local unitaries to dual-unitary gates. A closely related procedure works surprisingly well also for general (non-dual) unitaries, which allows for detailed numerical explorations of $K_D(U)$. We find $K_D(U)$ analytically for several special families of unitaries which also verifies the numerical procedure.

\section{General considerations and bounds for $K_D(U)$ } Let the operator Schmidt decomposition of an arbitrary two-qudit unitary gate $U$ be:
\begin{equation}\label{eq:Schmidt}
 U=\sum_{i=1}^{d^2} \sqrt{\lambda_i}\; m^A_{i} \otimes m^B_{i}.  
\end{equation}
Here the set $\{m^{A}_i, \; i=1,\cdots,d^2\}$ forms an orthonormal operator basis in subspace $A$ that is
\[ \tr(m^{A}_{i} m^{A\, \dagger}_j)=d\delta_{ij},\]
and $m^B$ is a similar set in subspace $B$. The operator Schmidt coefficients $\lambda_k$ are chosen to be in decreasing order, $1 \geq \lambda_1\geq \cdots \geq \lambda_{d^2} \geq 0$, and from the unitarity of $U$ it follows that $\sum_{k=1} ^{d^2} \lambda_k =1$.

The Schmidt decomposition provides measures of operator entanglement, for example one such is the linear entropy 
\begin{equation}\label{eq:E(U)}
    E(U)=1-\sum_{k=1}^{d^2}\lambda_k^2.
\end{equation}
This range is $0 \leq E(U) \leq E(S)=(d^2-1)/d^2$, and is $0$ iff $U$ is a product unitary (in which case $K_D(U)=0$).
The maximum value is attained when for all $k$, $\lambda_k=1/d^2$, and are the operator equivalents of Bell states. The swap operator $S$ ($S\ket{kl}=\ket{lk}$) achieves this value and is an important example of a maximally entangled unitary operator. However, $S$ is by far not the only unitary operator to achieve this value.  If an unitary $U$ is such that $E(U)=E(S)$, this may also be taken as the {\it definition} of dual-unitary operators.
Restriction of an unitary $U$ to this special set will be generically denoted as $U_{\text{dual}}$, that is $E(U_{\text{dual}})=1$. One of our goal is to calculate $K_D(U_{\text{dual}})$, but we first turn to general statements and bounds.

For the distance measure $K_D(U)$ in Eq.~(\ref{eq:kd_def}), for the rest of the paper we use the Hilbert-Schmidt metric.
Define
\begin{eqnarray}\label{eq:K_D*defn}
K_D^*(U):= \|U-m^A_1\otimes m^B_1\|=\sqrt{2d^2-2d^2\sqrt{\lambda_1}}.   
\end{eqnarray}
As the Schmidt decomposition already provides the nearest product operator (for a proof in the case of states, see \cite{Peresbook}),
this quantity is the distance to the nearest product operator, removing the unitarity constraint from the local operators in Eq.~(\ref{eq:kd_def}). As $K_D(U)$ is the distance to a more constrained set, the following lower-bound follows:
\begin{equation}
\label{eq:kDlowerbound}
K_D(U) \geq K_D^*(U).
\end{equation}

Alternatively, the definition in Eq.~(\ref{eq:kd_def}) implies
\beq
\label{eq:kDsquared}
\begin{split}
K_D^2(U) &=  \min_{u_A,u_B}\|U-u_A \otimes u_B\|^2 \\& = \min_{u_A,u_B} \left( 2\, d^2-2\text{Re}\left[\text{tr}(U^\dagger ( u_A \otimes u_B))\right]\right) \\
    & =\, 2\,d^2-\max_{u_A,u_B}2 \left|\text{tr}(U^\dagger (u_A \otimes u_B))\right|.
\end{split}
\eeq
The second equality follows from the fact that the phase can be absorbed by the local unitaries. 
Expand local unitaries $u_A$ and $u_B$ in the orthonormal bases from the Schmidt decomposition of $U$ in Eq.~(\ref{eq:Schmidt}), $u_A=\sum_{i=1}^{d^2} \alpha_i \, m^A_{i}, \;\; u_B=\sum_{i=1}^{d^2} \beta_i\, m^B_{i}$, where $\sum_{i}|\alpha_i|^2=\sum_i |\beta_i|^2=1$. This leads to $\left|\text{tr}(U^\dagger (u_A \otimes u_B))\right|=$
\begin{equation}
    \left|d^2\sum_{i=1}^{d^2} \sqrt{\lambda_i} \, \alpha_i\beta_i\right|  \leq d^2\sqrt{\lambda_1}\sum_{i=1}^{d^2}|\alpha_i| |\beta_i| \leq  d^2\sqrt{\lambda_1}.
\end{equation}
Here the first inequality holds as $\lambda_1$ is the largest Schmidt coefficient, and the second follows from the Cauchy-Schwarz inequality. Using this in Eq.~(\ref{eq:kDsquared}) we indeed obtain the lower-bound as $K_D^*(U)$ in Eq.~(\ref{eq:kDlowerbound})

It is possible to obtain an upper-bound as
\begin{equation}
\label{eq:kDupperbound}
K_D(U) \leq K_D^*(U)+\sqrt{2d^2-2 \|m_1^A\|_1 \|m_1^B\|_1},
\end{equation}
where $\|A\|_1=\tr\sqrt{A A^{\dagger}}$ is the trace norm. 
If $m^A_1 = \overline{u}_A \sqrt{m_1^{A\, \dagger} m^A_1}$ is its polar decomposition and similarly for $B$, $ \overline{u}_{A,B}$ are the nearest unitaries to $m^{A,B}_1$ \cite{KyFan_1955}. The upper bound is obtained as by definition the distance of $U$ to the product $\overline{u}_A\otimes \overline{u}_B$ cannot be less than $K_D(U)$:
\begin{equation}
\begin{split}
&K_D(U)  \leq \|U-\overline{u}_A\otimes \overline{u}_B\| \\& =\|U-m^A_1\otimes m^B_1+ m^A_1\otimes m^B_1-\overline{u}_A\otimes \overline{u}_B\|
\\ & \leq \|U-m^A_1\otimes m^B_1\|+\|m^A_1\otimes m^B_1-\overline{u}_A\otimes \overline{u}_B\|.
\end{split}
\end{equation}
The final step follows from the triangle inequality and the upper-bound is immediately obtained.

These lower and upper-bounds on $K_D(U)$ are dependent only on the 
principal Schmidt eigenvalue $\lambda_1$ and its corresponding Schmidt operators $m_1^{A,B}$.
These bounds are tight, and it will be seen that the lower-bound is quite good, and is, in any case, far from the average value of the distance (squared) from local unitaries, which is $2d^2$.  
For the most part, we will concentrate on the lower bound that is maximized when $\lambda_1$ is the minimum possible value $=1/d^2$.  This happens only when {\it all} the $\lambda_i$ are equal and $=1/d^2$, which implies that $U$ is necessarily maximally entangled, which is the same as dual-unitary.
Therefore $K_D^*(U)\leq K^*_D(U_{\text{dual}})=\sqrt{2d^2-2d}$. For convenience we define 
\begin{equation}\label{eq:K_D*}
K^*_D:= \max_{U \in \mathbb{U}(d^2)}K_D^*(U)=\sqrt{2d^2-2d}.
\end{equation}
Providing a proof for $d=2$ and evidence for $d>2$, we conjecture that 
\begin{equation}\label{eq:K_Dconjecture}
\max_{U \in \mathbb{U}(d^2)}K_D(U)=K_D^*.
\end{equation}


\section{$K_D(U)$ for the two-qubit case} 

 As $K_D(U)$ is a local-unitary invariant,
it is sufficient to consider the nonlocal part of the canonical decomposition \cite{Barbara_2001,Khaneja_2000}:
\begin{equation}\label{eq:Cartan}
    U=\exp[i(c_1 \, \sigma_1 \otimes \sigma_1+c_2 \, \sigma_2 \otimes \sigma_2+c_3 \,\sigma_3 \otimes \sigma_3)].
\end{equation}
As the products of the Pauli matrices in this expression commute we have: 
\begin{equation}\label{eq:2QubitSchmidt}
    U= \prod_{k=1}^3\exp(ic_k\, \sigma_k \otimes \sigma_k) =\sum_{k=0}^3 \sqrt{\mu_k} \, (e^{i\theta_k} \sigma_{k})\otimes \sigma_{k}.
\end{equation}
Here the $\sigma_0$ is the $2\time2$ identity matrix and  $\mu_k$ are functions of the $c_i$, for explicit expressions see \cite{Balki_2011}. This is itself a Schmidt decomposition and the $\lambda_k$ are found by 
arranging $\mu_k$ in decreasing order, in particular $\lambda_1=\text{max}\{\mu_0,\mu_1,\mu_2,\mu_3\}$.
This implies that for any 2-qubit unitary operator, there exists a Schmidt decomposition where both pairs of orthonormal bases are unitary as well. Hence, 
\begin{equation}
K_D(U)=K_D^*(U)=\sqrt{8(1- \sqrt{\lambda_1})},
\end{equation}. In this case both the lower and upper bounds coincide and are exact. 
It immediately follows that $K_D(U)\leq K_D^*=2$.  Therefore $\max_{U\in \mathbb{U}(4)}K_D(U)=K^*_D=2$. This is achieved {\em  only} for the case of dual-unitary gates. These are parameterized in the Cartan form by the one parameter family: $c_1=c_2=\pi/4$, and $c_3$ is arbitrary. We will find this to be true for higher dimensions through numerical investigations, as discussed later.
  \begin{figure}[h!]
            \centering
            \includegraphics[scale=0.35]{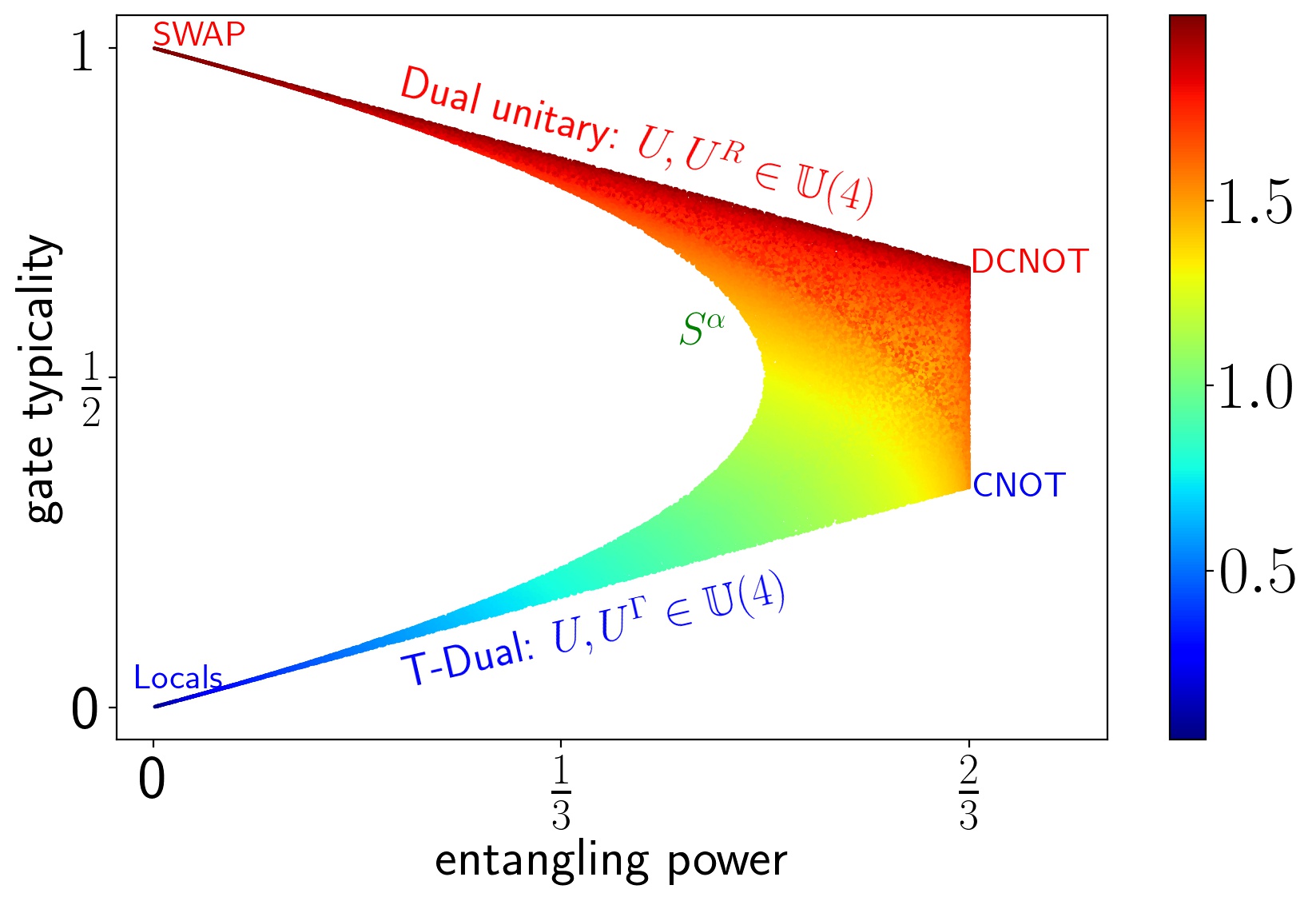}
            \qquad
            \caption{The distance to local gates, $K_D(U)$,  plotted as a color map in the space of entangling properties
            of the gate. We observe that the highest values are attained on the dual-unitary boundary line.},
            \label{fig:2qubit}
    \end{figure}
In Fig.~(\ref{fig:2qubit}) the $K_D(U)$ is shown for all possible two-qubit operators as a function of their entangling power $e_p(U)$ and gate-typicality $g_t(U)$ which are local unitary invariants \cite{Zanardi_2001,Bhargavi_2017,Bhargavi_2019}. 
For completeness, we recall their definitions.
The entangling power is the average linear entropy of $U\ket{\phi_A}\ket{\phi_B}$ when $\phi_{A,B}$ are uniformly (Haar) sampled from the subspaces and is related to the operator entanglement,  Eq.~(\ref{eq:E(U)}. Under appropriate scaling, we take 
    $e_p(U)=\left(E(U)+E(US)-E(S)\right)/E(S),$
where $S$ is the swap operator and $0 \leq e_p(U)\leq 1$. If $U$ is dual-unitary, $E(U)=E(S)$ and hence $e_p(U)=E(US)/E(S)$. This is maximum and is equal to $1$ iff $US$ is {\it also} dual-unitary. In general if $U $ is dual-unitary, $US$ is defined to be $\Gamma$-dual (or T-dual), as the partial transpose of the unitary is also unitary \cite{ASA_2021}. As the swap does not create any entanglement $e_p(S)=e_p(I)=0$, however $E(S)$ is the maximum possible, that is the swap is a very nonlocal gate. A complementary quantity, the so-called gate-typicality $g_t(U)$ is useful in separating the nonlocals from the swap and is defined by $g_t(U)=\left(E(U)-E(US)+E(S)\right)/(2 E(S))$. This vanishes for $U$ that is local or identity and is maximum $=1$ only for the swap.
For qubits, it is known that the maximum value of $e_p(U)$ is only $2/3$ and not $1$; for all other local dimensions the value of $e_p(U)=1$ is achieved.

Returning to Fig.~(\ref{fig:2qubit}), the dual unitary gates lie on the upper boundary line, while the $\Gamma$-dual make up the lower-boundary. The left boundary is the parabola containing all the fractional powers of swap, while the vertical right boundary connects the CNOT gate with its ``swap partner" obtained by multiplying it with the swap gate. It is interesting that the $K_D(U)$ strongly correlated with these invariants and the ``hottest" regions are in the vicinity of the dual-unitary upper-boundary and the ``cool" ones are in the vicinity of the $\Gamma$-dual and the locals. 

\section{$K_D(U)$ for dual unitary $U$ and arbitrary $d$}

Recall that dual-unitary operators have maximal operator entanglement and in their Schmidt decomposition
is flat with $\lambda_k=1/d^2$ for all $k$. It is also useful to think of them as retaining unitarity under 
a reshuffling of their elements. Define the realignment operation $R$ as
\begin{equation}\label{eq:Realign}
    \langle ij|U^R|kl \rangle=\langle ik|U|jl \rangle.
\end{equation}
The following may be easily verified: $(a\otimes b)^R=\ket{a} \bra{b^*}$, where $\bra{kk'}\ket{v}=\bra{k}v\ket{k'}$ is a vectorization of the matrix $v$. Thus taking the realignment of Eq.~(\ref{eq:Schmidt}) we get 
\begin{equation}\label{eq:URSchmidt}
    U^R=\sum_{k=1}^{d^2} \sqrt{\lambda_k} \ket{m_k^A}\bra{m_k^{B*}}. 
\end{equation}
If $\lambda_k=1/d^2$, it follows that $U^R U^{ R \dagger}=I_{d^2}$, and $U^R$ is unitary. Indeed $U$ is dual-unitary iff $U^R$ is also unitary \cite{KZ_2004,SAA_2020}. 

For dealing with dual-unitary operators it may be desirable to also deal with $U^R$ and it is not hard to see that $\left|\text{tr}(U^\dagger (u_A \otimes u_B))\right|=\left|\langle u_A|U^R|u_B^*\rangle \right|$.
Using this we can rewrite the expression for $K_D(U)$ in Eq.~(\ref{eq:kDsquared}) as 
\begin{equation}
\label{eq:kDsquaredAlt}
\begin{split}
K_D^2(U) 
=2d^2-2\max_{u_A,u_B} \left|\langle u_A|U^R|u_B^*\rangle \right|.
\end{split}
\end{equation}
Note that except in $K_D^{*}$, $*$ denotes complex conjugation. The vectorization of unitary operators such as $u_A$ lead to maximally entangled states $\ket{u_A}$, with normalization $\bra{u_A}\ket{u_A}=d$. In this approach, the problem of finding $K_D(U)$ is reduced to finding the maximum overlap of $U^R$ with maximally entangled states. We now state a property that helps with evaluating this and is true for $d=2$, but is a conjecture for $d>2$.

\subsection{UBB Conjecture} 

The conjecture is that for any unitary operator $V$ acting on two particle states of $d$-dimensions each, there exists at least one pair of maximally entangled state $|\Phi_0\rangle$ and $|\Phi_1\rangle$ such that, 
   \begin{equation}
       V|\Phi_0\rangle=|\Phi_1\rangle.
       \label{eq:ubell}
   \end{equation}
If this statement is true, then Eq.~(\ref{eq:kDsquaredAlt}) implies that for a dual unitary $U$, $K_D(U)=K^*_D=\sqrt{2 d^2-2d}$. This follows as in this case $U^R$ is also unitary and takes the place of $V$ in the conjecture, and the maximizing overlap, namely $d$, is achieved. To avoid confusing the unitary $U$ for which we are interested in finding $K_D$, we denote the generic unitary now as $V$.
This property is referred to as UBB for ``unitary-times Bell is Bell", and implies that the Schmidt decomposition of dual-unitaries can be chosen such that at least one pair $(m_i^A, m_i^B)$ can be unitary. 

To see this note that for dual-uniatries, as $U^R$ is also unitary, all the Schmidt values in Eq.~ (\ref{eq:URSchmidt}) are $\{ \lambda_k=1/d^2, \, 1\leq k \leq d^2\}$, and we may choose the set of Schmidt vectors $\{\ket{m_k^{B *}}/\sqrt{d},\, 1\leq k \leq d^2\}$ to be any orthonormal set such that the other set of Schmidt vectors is $\ket{m_k^{A}}=U^R \ket{m_k^{B *}}$. Thus if we choose say $\ket{m_1^{B *}}/\sqrt{d}$ to the one of the pair of UBB states corresponding to $U^R$, then $\ket{m_k^{A}}/\sqrt{d}$ will also be maximally entangled. This implies that $\ket{m_1^{B *}}$ and $\ket{m_1^{A}}$ are vectorizations of unitary operators as claimed. This renders the upper bound in Eq.~(\ref{eq:kDlowerbound}) also to be $K_D^*(U)=K_D^*$. 

Note that the property UBB is local-unitary invariant and that the canonical decomposition in the case of qubits, Eq.~(\ref{eq:Cartan}) is such that the standard Bell states are eigenvectors and the corresponding eigenvalues have absolute values of 1. The decomposition in Eq.~(\ref{eq:2QubitSchmidt}) also shows explicitly that all four Schmidt operators can be taken to be unitary. Thus for $d=2$, the UBB property is already proven valid in general. It is unsurprising that such a basic property as UBB has in fact been discussed earlier \cite{Puchala_2015}, however it is surprising that it remains, to the best of our knowledge, unproven (or contradicted) in general for $d>2$. 

\subsection{A special case where UBB is provable}

One important special case for which UBB is easily shown to hold for all local-dimensions $d$ is when $V=\mathcal{D}$ is diagonal, and hence also for all $V$ that is local-unitary connected to diagonals: $V=(u_A\otimes u_B) \mathcal{D} (u_A'\otimes u_B')$. If 
\begin{equation}
    \mathcal{D}=\text{diag}(e^{\phi_{11}}, \cdots e^{i \phi_{dd}}),
\end{equation}
this follows as $\mathcal{D}\ket{\Phi^+}=$
\begin{equation}
    \mathcal{D}\left(\frac{1}{\sqrt{d}}\sum_{k=1}^d\ket{kk}\right)= \ket{\Phi_1}=\frac{1}{\sqrt{d}}\sum_{k=1}^d e^{i \phi_{kk}}\ket{kk}.
\end{equation} 
That is, the standard maximally entangled state $\ket{\Phi^+}$ along with a phase-decorated state $\ket{\Phi_1}$, which is also maximally entangled, is one such required pair. In fact this can be generalized to provide a maximally entangled basis that remains maximally entangled under $\mathcal{D}$. Let 
\begin{equation}
    \ket{W_{mn}}=\frac{1}{\sqrt{d}} \sum_{\ell=1}^d e^{2 \pi i m \ell/d} \ket{\ell,\ell+n}. 
\end{equation}
This is a maximally entangled state and is the vectorization of $W_{mn}=T_1^m T_2^{-n}/\sqrt{d}$ where $T_1\ket{\ell}=e^{ 2\pi i \ell /d}\ket{\ell}$ and $T_2\ket{\ell}=\ket{\ell+1}$ are the clock-and-shift operators or generalized Pauli matrices. However, 
\begin{equation}
    \mathcal{D}\ket{W_{mn}}=\frac{1}{\sqrt{d}} \sum_{\ell=1}^d e^{2 \pi i m \ell/d} e^{i \phi_{\ell,\ell+n}} \ket{\ell,\ell+n}
\end{equation}
is also a maximally entangled state. This is the vectorization of $T_1^m T_2^{-n}\mathcal{D}_n/\sqrt{d}$, where $\mathcal{D}_n=\text{diag}\{e^{i \phi_{1,1+n}},\cdots, e^{i \phi_{d,d+n}}\}$. 

Note though that while $\mathcal{D}$ is not dual-unitary (it is $\Gamma$-dual), $S \mathcal{D}$ is, dual, where $S$ is the {\sc swap} operator. Such operators have been extensively used in the study of dual-unitary circuits as they provide a simple family \cite{Claeys_2021}, and in fact are self-dual: $(S\mathcal{D})^R=S\mathcal{D}$ \cite{ASA_2021}. The {\sc swap} $S$ itself is a dual-unitary operator and the UBB property clearly hold in this case, as any maximally entangled states $\ket{\Phi_0}$, remains maximally entangled.
Therefore we have at least $d^2$ orthonormal maximally entangled states $\ket{W_{mn}}$ that are such that 
$S \mathcal{D}\ket{W_{mn}}=\ket{W'_{mn}}$,
where 
\begin{equation}
    \ket{W'_{mn}}=\frac{1}{\sqrt{d}} \sum_{\ell=1}^d e^{2 \pi i m \ell/d} e^{i \phi_{\ell,\ell+n}} \ket{\ell+n,\ell}
\end{equation}
is also maximally entangled. A Schmidt decomposition exists with all $m_i^{A,B}$ being unitary. Thus we have proved that 
\begin{equation}
K_D(S\mathcal{D})=K^*_D 
\end{equation}
for any diagonal unitary $\mathcal{D}$. Note though that we have not evaluated $K_D(\mathcal{D})$ itself, later we present numerical results for these, while presenting analytical results for a special class of diagonal unitaries. 
More cases where UBB can be analytically shown are described below, but we now turn to a numerical procedure that gives us confidence that UBB is true. 

\section{Evidence for the UBB conjecture}
\subsection{The UBB algorithm}
Given a unitary operator $V$ that is $d^2$ dimensional, the following algorithm is used to find pairs of maximally entangled states that satisfy the UBB condition for a given bipartite gate. Pick a random $u_0 \in \mathbb{U}(d)$ from the CUE.
\begin{enumerate}
    \item Find $\ket{v_0'}=V\ket{u_0}$.
    \item From polar decomposition of $v_0'$, $v_0'=v_0 \sqrt{v_0{'\dagger}v_0'}$, 
    find the unitary $v_0 \in \mathbb{U}(d)$. Let this step be denoted $v_0=P[v_0']$.
    \item Find $\bra{u_1'}=\bra{v_0} V$.
    \item Evaluate the nearest unitary $u_1 \in \mathbb{U}(d)$ from $u_1=P[u_1']$.
    \item Iterate from Step 1, with $u_0$ being replaced with $u_1$.
\end{enumerate}
Note that the primed operators $v_0'$ and $u_1'$ are not unitary in general. The projection into the unitary space using polar decomposition, denoted $P$, yield the closest unitary matrices \cite{KyFan_1955}. This procedure iterated $n$ times yields the pair $(u_n,v_n)$ and as $n\rightarrow \infty$ their vectorization corresponds to a required pair $\ket{\Phi_0}=\ket{u_{\infty}}$ and $\ket{\Phi_1}=\ket{v_{\infty}}$, provided $v_n'$ and $u_n'$ themselves tend to unitary matrices. In this case $v_{\infty}=v'_{\infty}$ and $u_{\infty}=u'_{\infty}$. In cases where this is not satisfied, the seed $u_0$ is changed. We found without exception that this always verified the UBB property for arbitrary choices of $V$. 

\begin{figure}
    \centering
    \includegraphics[scale=0.55]{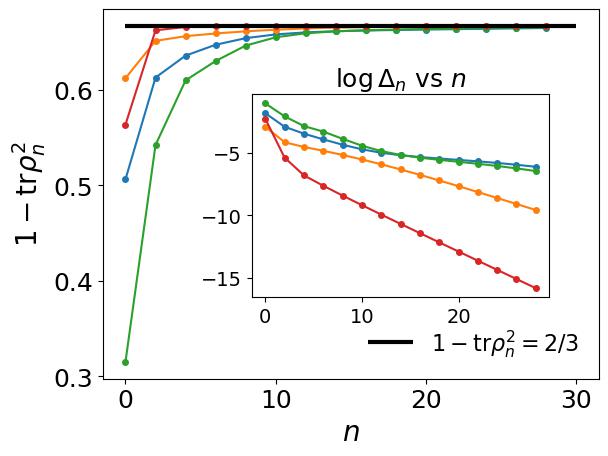}
    \caption{UBB demonstrated for a sample of random bipartite unitaries $V$ with $d=3$. The inset shows the approach to the maximum entropy $2/3$ via the difference $\Delta_n=2/3-(1-\tr\rho_n^2)$.}  
    \label{fig:ent_growth}
\end{figure}

\subsection{Convergence and entanglement growth under the algorithm}

That the algorithm converges for $V$ unitary is supported by the following observation: for a given maximally entangled seed state $\ket{u_0}$ define the distances, 
\begin{equation}
\begin{split}
    &d_0=\|V |u_0\rangle- |v_0\rangle\|_2, \; d_0'=\|V^\dagger |v_0\rangle- |u_1\rangle\|_2,\\ &  d_1 =\|V |u_1\rangle-|v_1\rangle\|_2,
    \end{split}
\end{equation}
where $|v_0\rangle,\,|u_1\rangle$ and $|v_1\rangle$ are maximally entangled (ME) states that minimise their respective expressions, and $\|\ket{v}\|_2=\sqrt{\braket{v}}$ is the vector Euclidean norm.  The distances satisfy $d_0 \geq d_1$ as 
\begin{equation}
\begin{split}
    &d_0=\min_{|\theta \rangle \in \text{ME}}\|V|u_0\rangle-|\theta\rangle\|_2 \\ & =\|V|u_0\rangle-|v_0\rangle\|_2=\| |u_0\rangle-V^\dagger|v_0\rangle\|_2 \\& 
    \geq \min_{|w \rangle \in \text{ME}} \||w\rangle-V^\dagger|v_0\rangle\|_2 =d_0'.
    \end{split}
\end{equation}
 Similarly $d_0' \geq d_1$, and hence the result follows. In general this shows that $\{ d_{n}=\|V \ket{u_n}-\ket{v_n}\|_2, \; n=0,1,\cdots\}$
    is a nonincreasing sequence. Implications of this for entanglement in the states which is our main quantity of interest is now studied.
    
Let  
\begin{equation}
    \rho_n^A=\tr_B(V\ket{u_n}\bra{u_n}V^{\dagger})/d,
\end{equation}
be the reduced density matrix at step $n$, ideally this must tend to be maximally mixed for 
$V\ket{u_n}$ approaching a maximally entangled state. Let $V\ket{u_n}$ be the vectorization of 
say $M_n$, then $\rho^A_n=M_n M_n^{\dagger}/d$. From $v_n$ being the closest unitary to $M_n$, the polar decomposition gives $M_n=v_n \sqrt{M_n M_n^{\dagger}}\,v_n= \sqrt{d \rho_n^A}\, v_n$. Thus
\begin{equation}
\begin{split}
    &d_n=\|V \ket{u_n}-\ket{v_n}\|_2=\|M_n-v_n\|\\&=\|\sqrt{d \rho_n^A} v_n-v_n\|=\|\sqrt{d \rho_n^A} -\mathbb{1}_d\|,
    \end{split}
\end{equation}
where the fact that the Hilbert-Schmidt norm of matrix versions of vectors coincides with their Euclidean norms is used. The non-increasing property of the distances $d_n$ then implies that 
$\|\sqrt{\rho_n^A}-\mathbb{1}_d/\sqrt{d}\|$ is also non-increasing, and hence that
the R\'enyi entropy of order $\frac{1}{2}$: $S_{1/2}=2 \log (\tr\sqrt{\rho_n^A})$, is non-decreasing.
These statements make it plausible that under this algorithm $\rho_n^A \rightarrow \mathbb{1}/d$ as $n \rightarrow \infty$.

Numerical results show that all R\'enyi entropies typically grow monotonically as the Fig.~(\ref{fig:ent_growth}) shows this for the case of the linear entropy $1-\tr(\rho_n^A)^2$. The approach of $\rho_n^A$ to being maximally mixed and hence $V\ket{u_n}$ being maximally entangled is shown for local dimension $d=3$, the smallest
dimension in which the UBB is a conjecture. However, starting from random seeds, for $d=3,4$ one step reduction of the linear entropy is observed in  about $10\%$ of cases and for $1\%$ of cases in two consecutive steps. Further evidence that the algorithm introduced is convergent comes from the fact that only fixed points are allowed cycles, see Appendix~\ref{app:fixedpoints}.
\begin{figure}
    \centering
    \includegraphics[scale=0.55]{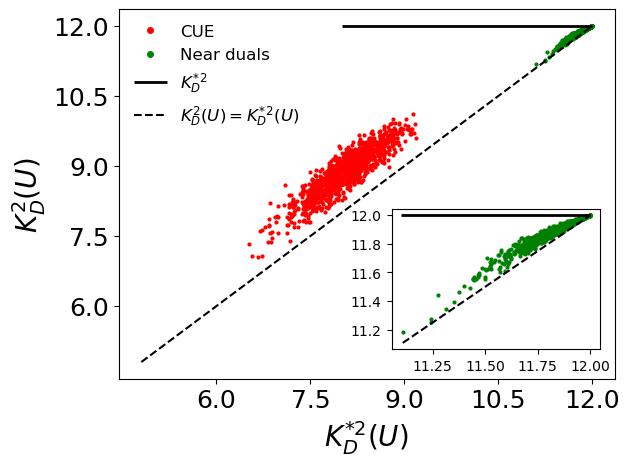}

    \caption{$K_D(U)$ obtained from the algorithm {\em vs} the corresponding analytical lower bound $K_D^*(U)$. This is displayed for random unitaries from CUE of size $d^2=9$, and random unitaries in the neighborhood of dual unitaries (also magnified in the inset),  number of samples taken in each case is $10^3$. The horizontal solid line is at $K_D^2(U)=K_D^{*2}=12$, and the dashed line is the equality $K_D^2(U)=K_D^{*2} $.}
    \label{fig:KD_KDS_plot}
\end{figure}
    

\section {$K_D(U)$ for general $U$ and $d$:} 

\subsection{Generalized algorithm}

When $U$ is dual unitary, with $V=U^R$, the UBB algorithm converges such that $u_{\infty}$, $v_{\infty}$, $u'_{\infty}$ and $v'_{\infty}$
are all unitary. For general $U$, not necessarily dual, with $V=U^R$, remarkably enough, the same algorithm still converges, although now $V$ is not in general unitary. Crucially, generally $u'_{\infty}$ and $v'_{\infty}$ are not unitary, and the overlap $|\braket{v_{\infty}}{v'_{\infty}}|=|\braket{u'_{\infty}}{u_{\infty}}|=|\bra{v_{\infty}}U^R\ket{u_{\infty}}|$ determines the $K_D(U)$. It is found that a variation of a few seeds $u_0$ typically yields the maximum value of this overlap, and hence the $K_D(U)$.
It was verified that for dual unitaries this algorithm yields $K_D(U)= \sqrt{2d^2-2d}$, as well as gives consistent results for the qubit case: $K_D(U)=K_D^*(U)$ and for the non-dual unitary cases in $d>2$ where exact evaluations are available, such as the cases discussed below.

Figure~(\ref{fig:KD_KDS_plot}) shows numerically converged values of $K_D(U)$ for a sample of random realizations of $U$ sampled from the CUE and random unitaries in the neighbourhood of dual unitaries for $d=3$. Crucially this algorithm allows us to verify that there were no $U$ such that $K_D(U)>K_D^*$. The $K_D(U)$ for random unitaries in the immediate neighborhood of dual unitaries (near duals in Fig.~(\ref{fig:KD_KDS_plot})) all have values less than $K_D^*$ as accentuated in the inset.  Thus extensive numerical calculations support what seems intuitive: the maximally entangled dual unitary operator set is farthest from product unitaries, and all of them are equally distant, with $K_D(U)=K_D^*$. This justifies the concentric rings in Fig.~(\ref{fig:cartoon}). Some partial analytical results in this context can be obtained, for example we show in the Appendix~\ref{app:block} that $K_D(U)<K_D^*$ if $U$ is block diagonal. 

\subsection{Case of diagonal unitaries}

Diagonal unitary matrices have been studied from the point of view of using them as approximate sources of 2-designs \cite{Nakata2017}, their entangling powers \cite{ArulKarolPuchala}, and as important constituents of quantum algorithms \cite{Welch2014} such as Grover's search. In the context of this paper, they provide a simpler set of matrices which are not dual-unitary, but share some properties with these. In particular there is a special subset that has maximal operator entanglement allowed for diagonal unitaries,
whose $K_D(U)$ can be analytically calculated with a generalized version of the UBB being explicitly valid. The special subset consists of diagonal unitaries of size $d^2$ constructed from a ($d \times d$) complex Hadamard matrix (CHM) \cite{Tadej_2006} which are unitary matrices filled with phases (up to normalization) 
\beq (u_{CHM})_{jk}=e^{i \phi_{jk}}/\sqrt{d}, \eeq 
with $\phi_{jk}$ all real. A well-known important example of a complex Hadamard matrix is the discrete Fourier transform:
$F_d=\sum_{j,k=1}^d e^{ 2 \pi i \, jk/d} \ket{j}\bra{k}/\sqrt{d}$, but the analysis below applies to any CHM.

For a given $d-$dimensional $u_{CHM}$, let
\begin{equation}
    \mathcal{D}_{H}=\sum_{j,k=1}^d e^{i \phi_{jk}}  \ket{jk}\bra{jk}
\end{equation}
be the corresponding diagonal unitary in $\mathcal{H}_d \otimes \mathcal{H}_d$. Then
\begin{equation}
    \mathcal{D}_H^R=\sum_{j,k=1}^d e^{i \phi_{jk}} \ket{jj} \bra{kk}=\sqrt{d}\, u_{CHM} \oplus 0_{d^2-d},
\end{equation}
where the $d^2$ dimensional space is split into a direct sum of the $d-$dimensional subspace $\{\ket{jj}, j=1, \cdots,d\}$ and the 
complementary $d^2-d$ dimensional space. It follows that $\mathcal{D}_{H}^R \mathcal{D}_{H}^{R \dagger} = d \, \mathbb{I}_{d} \oplus 0_{d^2-d}$, 
and hence, from Eq.~(\ref{eq:Schmidt}) and Eq.~(\ref{eq:URSchmidt}), the Schmidt decomposition of $\mathcal{D}_{H}$ is of rank $d$:
\begin{eqnarray}
\begin{split}
    & \mathcal{D}_{H}= \frac{1}{\sqrt{d}}\sum_{j=1}^d m_j^A \otimes m_j^B,  \\ & \mathcal{D}_{H}^R=\frac{1}{\sqrt{d}} \sum_{k=1}^{d} \ket{m_j^A}\bra{m_j^{B*}},
\end{split}
\end{eqnarray}
and all the non-zero Schmidt numbers are $\lambda_i=1/d$, $i=1, \cdots, d$.

\begin{figure}
    \centering
    \includegraphics[scale=0.52]{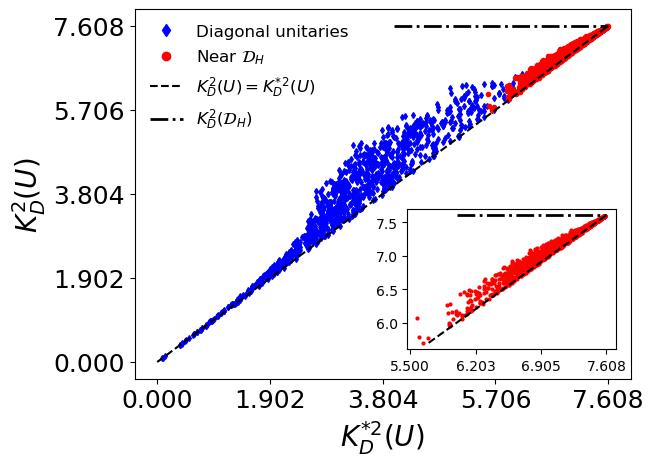}
    \caption{$K_D(U)$ obtained from the algorithm {\em vs} the corresponding analytical lower bound $K_D^*(U)$. This is displayed for random diagonal unitaries of size $d^2=9$, and random unitaries in the neighborhood of $\mathcal{D}_H$ (also magnified in the inset),  number of samples taken in each case is $10^3$. The horizontal solid line is at $K_D^{2}(\mathcal{D}_H)=18-9\sqrt{3}\approx 7.6 $, and the dashed line is the equality $K_D^2(U)=K_D^{*2} $.}
    \label{fig:KD_KDS_DH_plot}
\end{figure}

We now invoke the remarkable result \cite{Idel_2015} that for {\em any} unitary matrix $u$, there exists a pair of unimodular vectors  $\ket{e_1},\ket{e_2}$, consisting of pure phases, such that $u \ket{e_1}=\ket{e_2}$. Let $\ket{e_1},\ket{e_2}$ be such a pair for $u_{CHM}$. Define the $d^2$ dimensional vector $\ket{e_1'}$ by $\braket{jk}{e_1'}=0$ if $j\neq k$ and $\braket{jj}{e_1'}=\braket{j}{e_1}$,
and similarly define $\ket{e_2'}$. It follows that $\mathcal{D}^R_{H} \ket{e_1'}= \sqrt{d}\ket{e_2'}$. Thus we can take $\ket{m_1^{ B *}}=\ket{e_1'}$ and $\ket{m_1^A}=\ket{e_2'}$. The other $d-1$ orthogonal vectors $\ket{m_j^{B*}}$, $j>1$, can be taken to be orthogonal to $\ket{e_1'}$ and the $\ket{m_j^A}$ are defined by the action of $\mathcal{D}^R_{H}$ on them. Note that $m_1^A$ and $m_2^B$ are unitary as they
are diagonal and have only phases in them. Thus in the Schmidt decomposition of $\mathcal{D}_{H}$ at least one pair $(m_i^A, m_i^B)$ may be chosen unitary. Hence for this family of diagonal unitaries the lower bound for $K_D(U)$ in Eq.~(\ref{eq:kDlowerbound}) is tight,  and 
\beq
K_D(\mathcal{D}_{H})=\sqrt{2d^2-2d\sqrt{d}}\eeq is strictly less than the distance of dual unitaries from nearest products, namely $K^*_D$, in Eq.~(\ref{eq:K_D*}). 

In Fig.~(\ref{fig:KD_KDS_DH_plot}), $K^2_D(U)$ is plotted for a random ensemble of diagonals $\mathcal{D}$ and perturbed $\mathcal{D}_H$ diagonal unitaries for $d=3$. The perturbed $\mathcal{D}_H$ diagonal unitaries are obtained by multiplying the diagonal unitary obtained from a complex Hadamard matrix (for example, Fourier matrix) with diagonal unitaries that are close to the identity matrix.
For small values of $K_D^*(U)$, there is a one-to-one mapping between $K_D^*(U)$ and $K_D(U)$ as the largest Schmidt value almost entirely determines the non-local properties. For large values of $K_D(U)$, there is a correlation with $K_D^*(U)$, but also the numerical results point to the role of other local unitary invariants as well. The value of $K^2_D(\mathcal{D}_H)$ in this case is $\approx 7.6$ and we notice that all numerical values
are less than this, making it possible that $K_D(\mathcal{D})\leq K_D(\mathcal{D}_H)$, that is the CHM based diagonal unitaries have
the largest value of $K_D$ within the subset of diagonal unitaries. Indeed they have the largest possible operator entanglement $E(U)$,
for a study of the operator entanglement in diagonal unitaries, see \cite{ArulKarolPuchala}.

\subsection{$K_D(U)$ for fractional powers of swap}

Another family of unitary gates in $d \geq 2$ for which the lower-bound $K_D^*(U)$ is tight is that of fractional powers of {\sc swap} defined as $S^\alpha=\cos(\pi \alpha/2)\,\mathbb{I}_{d^2}+i\sin(\pi\alpha/2)\,S; \, \alpha \in [0,1]$.
Unlike the previous example, $(S^{\alpha})^R$ is full-rank (except for $\alpha=0$, with the largest singular value being distinct and the others are equal. As the largest singular value corresponds to the maximally entangled standard Bell state, we get
that the lower bound is still exact and 
\beq
K_D(S^{\alpha})=\sqrt{2 d^2-2 d \sqrt{d^2 \cos^2(\pi \alpha/2)+\sin^2(\pi \alpha/2)}},
\eeq
which interpolates between $0$, when $\alpha=0$, corresponding to the identity operator,
to the maximum $K^*_D$ when $\alpha=1$ and the {\sc swap} gate is dual unitary. Although in the examples above $K_D(U)=K_D^*(U)$, it is important to note that such an equality does not hold in general. This is shown for a class of unitaries in the Appendix~\ref{app:CZ} where $K_D(U)$ can be calculated analytically and is strictly larger than $K_D^*(U)$.

\section{Summary and outlook}

The distance of a general bipartite unitary on symmetric subspaces to product unitaries has 
been considered here. This has been considered as resource from early on. Here we relate this to dual-unitary operators showing in $d=2$ that they maximize the distance to nearest product unitaries. In $d>3$, we provide strong numerical evidence that this continues to hold. This is equivalent to the 
proving a conjecture that every bipartite unitary preserves the entanglement of a maximally entangled state. We have presented a simple Sinkhorn-like algorithm for finding such states for any bipartite unitary.

This is referred to as UBB in this paper, but was noted earlier in the literature \cite{Puchala_2015} where ``mutually entangled states" were considered. It is worth mentioning that this property has been discussed therein from a geometric perspective of non-displacable manifolds.
A recent work \cite{Lie_Kim2023} has also found use for the UBB property 
in the construction of unital generalized transpositions. 

A generalized version of the UBB, wherein the input matrix is the realignment of unitary matrices is also seen to exist wherein a pair of maximally entangled states and non-maximally entangled states form a periodic orbit under a algorithm. The exploration of the existence of limit sets of these algorithms is of interest as they in general seem to converge. It indicates that a generalized version of UBB holds whenever the operator (not necessarily unitary) is maximally entangled or, equivalently, it's realignment is unitary.

The case of diagonal unitary matrices presents a proxy for the larger group, however even in this limited subset the $K_D(U)$ problem is not fully resolved. The diagonals constructed from complex Hadamard matrices forms the equivalent of the general dual-unitary matrices. In this case we have been able to explicitly evaluate the distance to the nearest product unitary, thanks to a remarkable result concerning a Sinkhorn-like decomposition of a unitary matrix. However, even here it is left unresolved that these maximize the distance to the nearest 
product unitaries, although again there is excellent numerical evidence that such is the case.

Possible future directions, apart from proving the UBB property, include the behaviour of $K_D(U)$ for other norms, and
the preferred role of dual unitaries found here.
It can be shown that $K_D(U)$ over the Hilbert-Schmidt distance is not smaller than $K_D(U)$ over other Schatten-p norms, $p \geq 2$, making it an useful bound on circuit complexity. The case of multipartite unitary gates is of natural further interest.

\begin{acknowledgments}
We are grateful to Karol {\.Z}yczkowski for inputs concerning the UBB property and pointing to Ref.\cite{Puchala_2015}, and to
S. Aravinda for crucial discussions on the qubit case. Funding support from the Department of Science and Technology, Govt. of India, under Grant
No. DST/ICPS/QuST/Theme-3/2019/Q69 and MPhasis for supporting CQuiCC are gratefully acknowledged.
\end{acknowledgments}




\appendix
\section{Stability of $K_D(U)$ using the Hilbert-Schmidt norm}
\label{app:stability}
We prove the claim that the normalised Hilbert-Schmidt strength measure is a stable function, i.e, $f(U \otimes \mathbb{I})=f(U)$. 
For simplicity, we show stability by adding a single ancillary party to a bipartite unitary gate, but the extension to multipartite gates is trivial. If $U$ is a unitary operator of dimension $d^2 \times d^2$ and $\mathbb{I}_{d'}$ is an ancillary Identity operator of dimension $d'$, we have that $K_D^2(U \otimes \mathbb{I}_{d'})$
\begin{equation}
   =\min_{\substack{\text{$u_A,u_B \in \mathbb{U}(d)$} \\ \text{$u_C \in \mathbb{U}(d')$}}} 2d^2d'-2\text{Re}\left[\tr(U^\dagger(u_A\otimes u_B \otimes u_C))\right],
\end{equation}
Since we are interested in maximising the function in the expression on the right, we can add a phase to the local unitaries appropriately and make the expression real. Thus, $K_D^2(U \otimes \mathbb{I}_{d'})$
\begin{align*}
&=2d^2d'-2\max_{u_A,u_B,u_C} \left \vert \tr\left[(U^\dagger \otimes \mathbb{I})(u_A\otimes u_B \otimes u_C) \right] \right \vert\\
    &=2d^2d'-2\max_{u_A,u_B,u_C}\left \vert \tr(U^\dagger(u_A\otimes u_B))\tr(u_C) \right \vert \\
    &=2d^2d'-2d'\max_{u_A,u_B} \left |\tr(U^\dagger(u_A \otimes u_B)) \right |
\end{align*}

This follows from the inequality, $|\tr(u)| = |\tr(\mathbb{I}u)|\leq \tr|\mathbb{I}|=d$, where $u \in \mathbb{U}(d)$ and equality holds when $u = \mathbb{I}_d$. Therefore, $K_D^2(U \otimes \mathbb{I}_{d'})$
\begin{align*}
    &=d' \left ( 2d^2-2\max_{u_A,u_B} \left |\tr\left (U^\dagger \left (u_A\otimes u_B \right) \right ) \right| \right) \\
    &=d'K_D^2(U).
\end{align*}
By appropriate normalisation, $K_D'(U)=K_D(U)/\sqrt{\text{dim}(U)}$, the resultant function will be a stable strength measure.


\section{Fixed Points in the UBB Map}
\label{app:fixedpoints}
In this section, we show that if the UBB algorithm discussed in the main text converges to a cycle, the cycle is a fixed point.
Let us assume that the map has reached a period-$n$ cycle for some $n$. Each step has a corresponding projective distance and if this sequence decreases at any step of the cycle, it must increase at some stage of the cycle which follows from the periodicity of the cycle. As shown in the main text, these projective distances are non-increasing. Thus, in any period cycle the projective distances are all equal.

Now, let us say that the map begins from the state that corresponds to the first step of a period cycle. Then, 
\begin{equation}
\begin{split}
    d_0&=\min_{\ket{\theta} \in \text{ME}} \|U\ket{\phi_0}-|\theta\rangle\|^2\\& =\|U|\phi_0\rangle-|\phi_1\rangle\|^2.
    \end{split}
\end{equation}
However, 
\begin{equation}
\begin{split}
    &  \|U|\phi_0\rangle-|\phi_1\rangle\|^2=\||\phi_0\rangle-U^\dagger|\phi_1\rangle\|^2  \\
    &  \geq \min_{|w\rangle \in \text{ME}}\|U^\dagger|\phi_1\rangle-|w\rangle\|^2=d_1=\| |\phi_2\rangle-U^\dagger|\phi_1\rangle\|^2
\end{split}
\end{equation}
On expanding this expression, we see that this is equivalent to
\begin{equation*}
    |\langle \phi_0|U^\dagger|\phi_1\rangle|=\max_{|w\rangle \in \text{ME}}|\langle w|U^\dagger|\phi_1\rangle|
\end{equation*}
It can be shown that there exists an unique maximally entangled state obtained from the unitary part of the polar decomposition that maximises this expression \cite{KyFan_1955}. This means that in order for $d_0=d_1$,  $|\phi_0\rangle$ minimises the R.H.S of the above equation and therefore, $\ket{\phi_2}=\ket{\phi_0}$. This implies that the map has reached a fixed point. In short, we show that if there exists a period-cycle in this map, it is a fixed point.

\section{Analytical considerations of $K_D(U)$ for block-diagonal unitaries}

\subsection{Upper-bound of $K_D(U)$ for block-diagonal unitaries}
\label{app:block}
We illustrate a proof that $K_D(U) \leq \sqrt{2d^2-2d}=K_D^*$ for the special family of block diagonal unitaries. Let $U$ be a block diagonal consisting of $d \times d$ unitary blocks $u_i$, $i=1,\cdots,d$.
\begin{equation*}
\begin{split}
     K_D^2(U) & =2d^2-2\max_{u_A,u_B}\left|\tr(U^{\dagger}(u_A \otimes u_B))\right|,  \\
    & =2d^2-2\max_{u_A,u_B}\left|\tr_A(\tr_B(U^{\dagger}\mathbb{I}_d \otimes u_B)u_A)\right|, \\
    & =2d^2-2\max_{u_A,u_B}| M \,u_A|, 
\end{split}
\end{equation*}
where $M$ is a $d \times d$ diagonal whose $i^{th}$ diagonal element is tr$(u_i^\dagger u_B)$. 
Using the inequality $\max_{u}|\tr( A \,u)|=\| A \|_1$, 
\begin{equation*}
  K_D^2(U)  =2d^2-2\max_{u_B}\sum_{i=1}^d{\left|\tr(u_i^\dagger u_B)\right|}.
\end{equation*}
From here, it is clear that for all $v \in \mathbb{U}(d)$,
\begin{equation*}
    K_D^2(U) \leq 2d^2-2\sum_{i=1}^{d}{\left|\tr(u_i^\dagger v)\right|}.
\end{equation*}
Let $v=u_1$, the first block in $U$ for instance, then
\begin{equation*}
    K_D^2(U) \leq 2d^2-2d-2\sum_{i=2}^{d}{\left|\tr(u_i^\dagger v)\right|} \leq 2d^2 -2d,
\end{equation*}
which concludes our proof. In fact, it can be shown that $K_D^2(U) <  2d^2 -2d$, which is consistent with the fact that no block diagonal unitary is dual \cite{ASA_2021}. 

\subsection{$K_D(U)$ for an nearly Identity class of block-diagonal unitaries}
\label{app:CZ}
In this subsection, we discuss a simple class of block diagonal unitaries which are Identity, except for one 
of the elements, which is changed to $-1$. Let 
\[U_{CZ}=\mathbb{I}_{d^2-1} \oplus -1 = \mathbb{I}_{d^2} - 2 \ket{d^2}  \bra{d^2}.\]
Note that for $d=2$, $U_{CZ}$ it is the well known controlled Z (CZ) gate. For $U_{CZ}$
\begin{equation*}
   K_D^{2}(U_{CZ})=2d^2 -\max_{u \in \mathbb{U}(d)} \sum_{i=0}^{d-1} (d-1)|\tr(u)|+|\tr(Mu)|
\end{equation*}
Where $M=\mathbb{I}_d-2\ket{d}\bra{d}$ is the non-trivial (last) block in $U_{CZ}$.

The maximization in this case is analytically tractable and it can be shown that 
\begin{equation}
    K_D(U_{CZ})=
    \begin{cases}
    2\sqrt{2-\sqrt{2}} \approx 1.53\;\; \text{for} \;\; d=2 ,\\
    
    \sqrt{18-10\sqrt{2}} \approx 1.96 \;\; \text{for} \;\; d=3,\\
    
    2 \; \text{for} \;\; d \geq 4.
    \end{cases}
\end{equation}

Note that for $d>2$, $K_D(U_{CZ})>K_D^*(U_{CZ})$ and the lower bound for $K_D(U)$ is not tight; $K_D(U)>K_D^*(U)$ for all $d>2$.
It is remarkable that $K_D(U)$ becomes independent of $d$ for $d>3$.
In fact it can be shown that $K_D^*(U)= 2-2/d +O(1/d^2)$, and approaches $K_D(U)$ for large $d$.

   
   

\end{document}